\documentclass[12pt]{elsart}

\usepackage{graphicx}
\usepackage{subfigure}
\usepackage{amsmath}
\usepackage{cite}
\usepackage{endfloat}

\journal{Physics Letters A}

\begin{document}

\begin{frontmatter}

\title{Phase Diagram of a Diluted Triangular Lattice Ising Antiferromagnet in a Field}
\author{M. \v{Z}ukovi\v{c}\corauthref{cor}},
\ead{milan.zukovic@upjs.sk}
\author{M. Borovsk\'{y}},
\ead{borovsky.michal@gmail.com}
\author{A. Bob\'ak}
\ead{andrej.bobak@upjs.sk}
\address{Department of Theoretical Physics and Astrophysics, Faculty of Science,\\ 
P. J. \v{S}af\'arik University, Park Angelinum 9, 041 54 Ko\v{s}ice, Slovak Republic}
\corauth[cor]{Corresponding author.}

\begin{abstract}
\hspace*{5mm} Magnetization processes and phase transitions in a geometrically frustrated triangular lattice Ising antiferromagnet in the presence of an external magnetic field and a random site dilution are studied by the use of an effective-field theory with correlations. We find that the interplay between the applied field and the frustration-relieving dilution results in peculiar phase diagrams in the temperature-field-dilution parameter space. 
\end{abstract}


\end{frontmatter}

\section{Introduction}
\hspace*{5mm} The Ising model with nearest-neighbor antiferromagnetic interactions on a triangular lattice is fully frustrated due to local geometric constraints that prevent simultaneous minimization of all the pairwise interactions. It has been solved exactly and found to display no ordering at any finite temperatures \cite{wannier,hout}. Only at $T=0$ K the model displays a critical phase with algebraically decaying correlations \cite{steph}. The ground state is highly degenerated and thus even a small perturbation can cause that the system will choose an ordered state. Such a perturbation can be an external magnetic field, which by several studies has been shown to produce a line of second-order phase transitions for a certain range of the field values \cite{metcalf,schick,netz}. At the transition the system passes from the ferrimagnetic phase with two sublattices aligned parallel and one antiparallel to the field at lower temperatures ($\downarrow\uparrow\uparrow$) to the paramagnetic phase in which all spins are aligned parallel to the field at higher temperatures ($\uparrow\uparrow\uparrow$). As a result of the frustration, at low temperatures the magnetization versus field curves display a broad plateau with the value of $m=1/3$. Similar stepwise plateaus with different magnetization values have also been observed in some other lattices, such as the recursive lattice \cite{anan}, the zig-zag ladder \cite{hovh07,hovh08} and even on the cluster level \cite{koch06,koch08}. However, besides the magnetic field, the degeneracy of the system can also be removed by considering the interactions of more distant neighbors \cite{glos,taka,rast,kors}.\\
\hspace*{5mm} Diluting the system with nonmagnetic impurities in zero field locally relieves frustration and can lead to a spin-glass order \cite{grest,ander}, although this scenario remains controversial for the current 2D system \cite{blac}. Spin-glass ordering has also been observed in another highly frustrated Ising antiferromagnet on a fcc lattice for a certain range of the magnetic concentration \cite{weng}. In Ref. \cite{moes} the effects of doping the triangular Ising antiferromagnet with slow moving holes has been studied. Besides the site-dilution problem, also a bond-dilution problem \cite{ding} and a random-bond Ising model  \cite{loh} on a triangular lattice have been investigated. In contrast to the uniform dilution, if only one sublattice of the triangular lattice is diluted then a long-range order can develop in the remaining two sublattices already at relatively low concentrations of the impurities \cite{kaya}. Very recently, Yao \cite{yao} has studied how dilution can modulate the frustration effect in the system in a field. For example, he found that even a small dilution causes that the broad frustration-induced $1/3$ magnetization plateau splits into a stepwise curve. The importance of such studies is amplified by the fact that there are some corresponding real magnetic compounds, such as $\mathrm{Ca_3Co_2O_6}$, that have been experimentally found to display some unconventional frustration-induced features \cite{hardy} and the current model has been successfully employed in their explanation \cite{kuda,soto}. \\
\hspace*{5mm} Motivated by the previous interesting findings, the objective of the present work is to systematically study effects of the uniform site dilution on the phase transitions in the triangular lattice Ising antiferromagnet in a field. 

\section{Model and effective-field approach}
\begin{figure}[t!]
\centering
    \includegraphics[scale=0.4]{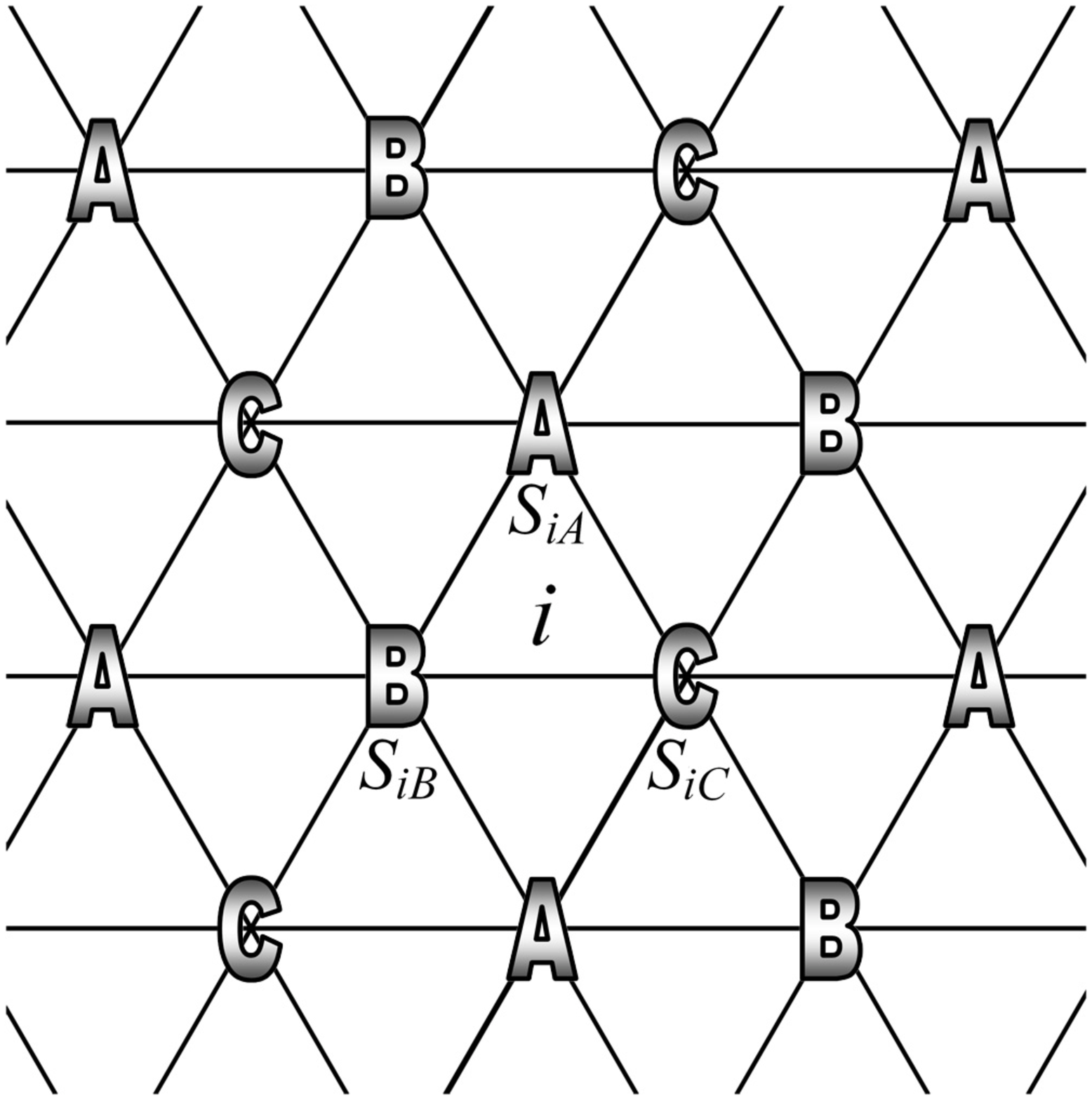}
\caption{Triangular lattice partition into three sublattices A,B and C.}
\label{fig:lattice}
\end{figure}

\hspace*{5mm} The model Hamiltonian is given by
\begin{equation}
\label{Hamiltonian}
H=-J\sum_{<i,j>}\xi_{i}\xi_{j}S_{i}S_{j}-h\sum_{i}\xi_{i}S_{i},
\end{equation}
where $S_{i}=\pm1$, are the Ising spin variables, $h$ is the external magnetic field, $J<0$ is the exchange interaction constant, and $<i,j>$ is the sum extending over all nearest neighbor (NN) pairs. $\xi_{i}$ are quenched, uncorrelated random variables chosen
to be equal to $1$ with probability $p$, when the site $i$ is occupied by a magnetic atom and $0$, with probability $1-p$ otherwise. Then the probability distribution is given by $P(\xi_i)= p\delta(\xi_i-1)+(1-p)\delta(\xi_i)$ and $p$ represents the mean concentration of magnetic sites. \\
\hspace*{5mm} We employ an effective-field theory (EFT) with correlations (see e.g., \cite{kane}), based on a single-site cluster approximation with the attention focused on a cluster comprising just a single spin, labeled $i$, and the NN spins with which it directly interacts. Following Ref. \cite{balc} we get the exact relation
\begin{equation}
\label{callen}
\xi_{i}\langle S_{i}\rangle= \xi_{i}\langle\tanh[\beta(J\sum_{j=1}^{z}\xi_{j}S_{j}+h)]\rangle,
\end{equation}
where $z$ is the number of NNs of the site $i$ (i.e., the coordination number), $\beta=1/k_{B}T$, and $\langle\ldots\rangle$ denotes a thermal average for a fixed spatial configuration of the occupied sites. Applying the differential operator technique \cite{honm} to the identity (\ref{callen}) and using the exact relations
\begin{equation}
\label{ex1}
\exp(\lambda \xi_{j})=\xi_{j}\exp(\lambda)+1-\xi_{j},
\end{equation}
\begin{equation}
\label{ex2}
\exp(\mu S_{j})=\cosh(\mu)+S_{j}\sinh(\mu),
\end{equation}
one obtains
\begin{equation}
\label{conf}
\xi_{i}\langle S_{i}\rangle=\xi_{i}\Big\langle
\prod_{j=1}^{z}[\xi_{j}\cosh(\beta JD)+\xi_{j}S_{j}\sinh(\beta JD)+1-\xi_{j}]\Big\rangle\tanh(x+\beta h)|_{x=0},
\end{equation}
where $D=\partial/\partial x$ is the differential operator.
In order to carry out the configurational averaging over the occupational numbers $\xi_{i}$,
let us assume that NNs of the site $i$ are completely independent of each other by
taking an approximation
\begin{equation}
\label{approx}
\langle S_{j}S_{k}\ldots S_{l}\rangle \approx\langle S_{j}\rangle\langle S_{k}\rangle\ldots\langle S_{l}\rangle.
\end{equation}
In spite of this simplification, we note that this approximation is quite superior to the standard mean-field theory,
since here, by using Van der Waerden identity (\ref{ex2}), the relations like $S_{j}^{2}=1$ are exactly taken into account. \\
\hspace*{5mm} However, unlike on some other lattices, on a triangular lattice NNs of the central spin $i$ include pairs of spins that are also mutual NNs and, therefore, their decoupling by the approximation (\ref{approx}) could result in rather high inaccuracies. Even more importantly, such a straightforward application of EFT would lead to a complete lost of the frustration and therefore inevitably incorrect results. In order to include all the NN interactions and the effect of the geometrical frustration, we chose to partition the lattice into three interpenetrating sublattices A, B and C in such a way that spins on one sublattice only interact with spins from the other two sublattices (see Fig.~\ref{fig:lattice}). Then all the NN interactions are accounted for and the frustration arises from the effort to simultaneously satisfy all the mutual intersublattice interactions, which are antiferromagnetic and isotropic i.e., $J_{\mathrm{AB}}=J_{\mathrm{AC}}=J_{\mathrm{BC}}\equiv J<0$. As an example, let us consider the situation in a selected triangular plaquette $i$ in Fig.~\ref{fig:lattice}. If the spin $S_{i\mathrm{A}}$ is in the state $+1$ then the energy is minimized if all its NNs, including $S_{i\mathrm{B}}$ and $S_{i\mathrm{C}}$, are in the state $-1$. However, for the spin $S_{i\mathrm{B}}$ in the state $-1$ it would be energetically favorable if all {\em its} NNs, including $S_{i\mathrm{A}}$ and $S_{i\mathrm{C}}$, were in the state $+1$, which creates frustration for the spin $S_{i\mathrm{C}}$. Given the lattice partition, the expression (\ref{conf}) can be written for individual sublattices in the form
\begin{align}
\label{conf_sub}
\xi_{i}\langle S_{i}\rangle=&\xi_{i}\Big\langle
\prod_{j=1}^{z_{1}}[\xi_{j}\cosh(\beta JD)+\xi_{j}S_{j}\sinh(\beta JD)+1-\xi_{j}]\nonumber \\
&\times\prod_{k=1}^{z_{2}}[\xi_{k}\cosh(\beta JD)+\xi_{k}S_{k}\sinh(\beta JD)+1-\xi_{k}]\Big\rangle\tanh(x+\beta h)|_{x=0},
\end{align}
where $z_{1},z_{2}$ are the numbers of NNs of the spin $S_{i}$ on a given sublattice that belong to the remaining two sublattices. Then, by performing configurational averaging for all three sublattices, the respective sublattice magnetizations $m_{\mathrm{X}}=\langle\xi_{i}\langle S_{i\mathrm{X}}\rangle\rangle_{c}$, X=A,B and C, can be calculated from the set of coupled equations
\begin{equation}
\begin{array}{l}
		\label{sub_mag}
		m_{\mathrm{A}} = p \left( a + b m_{\mathrm{B}} \right)^3 \left( a + b m_{\mathrm{C}} \right)^3 \tanh \left( x + \beta h \right) |_{x=0}, \\
				m_{\mathrm{B}} = p \left( a + b m_{\mathrm{A}} \right)^3 \left( a + b m_{\mathrm{C}} \right)^3 \tanh \left( x + \beta h \right) |_{x=0}, \\
				m_{\mathrm{C}} = p \left( a + b m_{\mathrm{A}} \right)^3 \left( a + b m_{\mathrm{B}} \right)^3 \tanh \left( x + \beta h \right) |_{x=0},
\end{array}
\end{equation}
where $a = 1 - p + p \cosh \left( \beta J D \right)$, $b = \sinh \left( \beta J D \right)$ and $p=\langle\xi_{i}\rangle_{c}$ is the same for all three sublattices. The explicit form of Eqs. (\ref{sub_mag}) can be calculated by using the mathematical relation $\exp(\alpha D)f(x)=f(x+\alpha)$. Then we can define the total magnetization per site $m=(m_{\mathrm{A}}+m_{\mathrm{B}}+m_{\mathrm{C}})/3$ and the order parameter $o= [\mathrm{max}(m_{\mathrm{A}},m_{\mathrm{B}},m_{\mathrm{C}})-\mathrm{min}(m_{\mathrm{A}},m_{\mathrm{B}},m_{\mathrm{C}})]/2$. The latter serves to localize phase boundaries between the ferrimagnetic ($\downarrow\uparrow\uparrow$) and the paramagnetic ($\uparrow\uparrow\uparrow$) phases.

\section{Results and discussion}
\begin{figure}
\centering
    \includegraphics[scale=0.5]{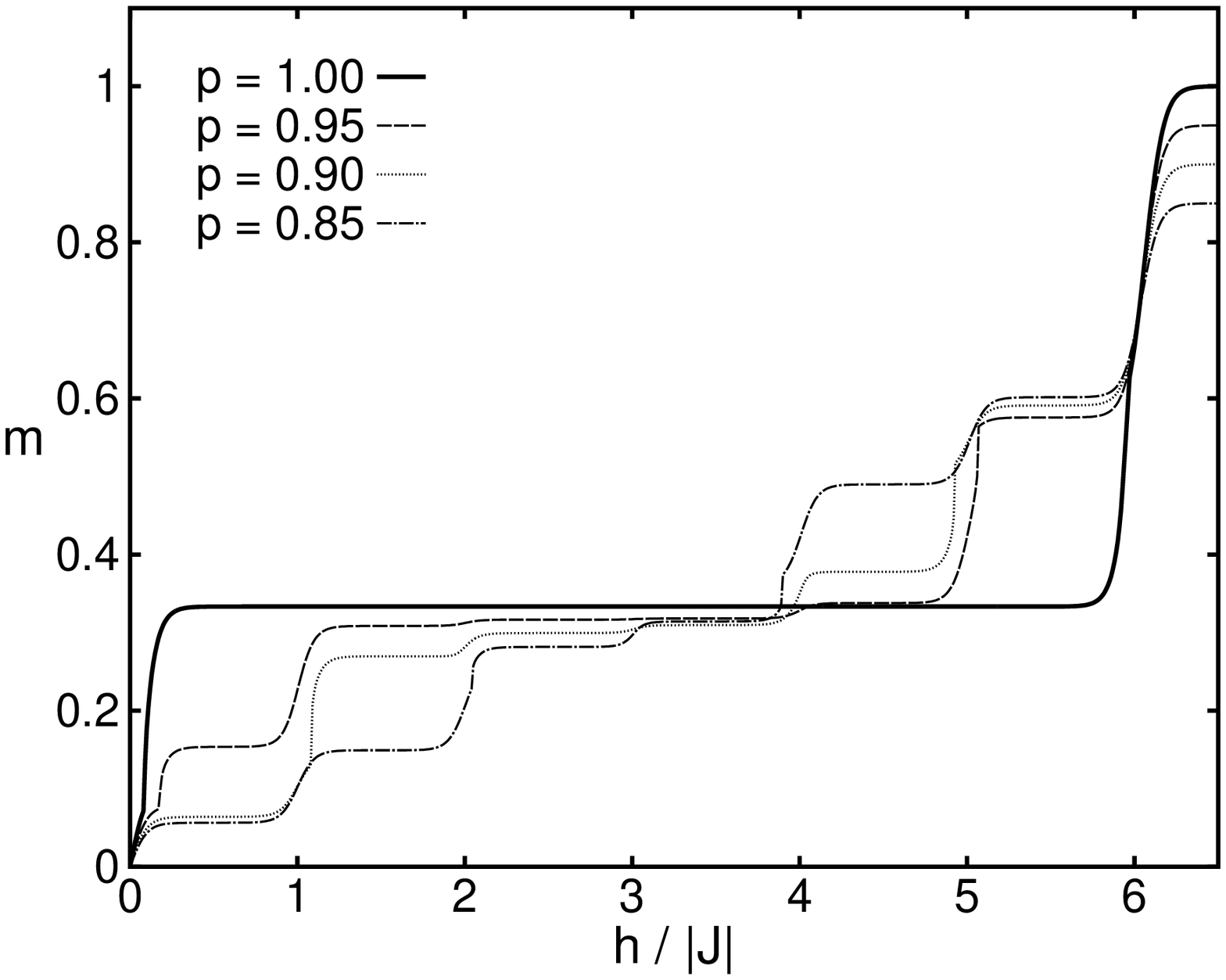}
\caption{Magnetization vs magnetic field plots for different values of the concentration $p$ at $k_BT/|J|=0.1$.}
\label{fig:m-h}
\end{figure}

\hspace*{5mm} In zero field, Eqs. (\ref{sub_mag}) have only trivial solution at all temperatures, which means that our effective-field approach reproduces the exact result of no long-range order down to $T=0$ K \cite{wannier,hout}. In Fig.~\ref{fig:m-h} we plot the field dependence of the magnetization at low temperatures for the pure case $(p=1)$ as well as diluted cases ($p<1$) with different degrees of dilution. For the pure case, the curve displays the typical frustration-induced broad $1/3$ plateau, in accordance with some previous theoretical \cite{yao,HuDu08} and experimental \cite{hardy} observations. When the system is diluted, we can observe formation of multiple steps at about integer values of the field $h/|J|=1,2,3,4,5$, which get more pronounced as the dilution increases. This behavior is in agreement with the Monte Carlo results \cite{yao} and it could be attributed to a gradual flipping of some dilution-relieved spins to the field direction when the Zeeman term contribution overcomes their exchange energy.

\begin{figure}[t!]
\centering
    \includegraphics[scale=0.5]{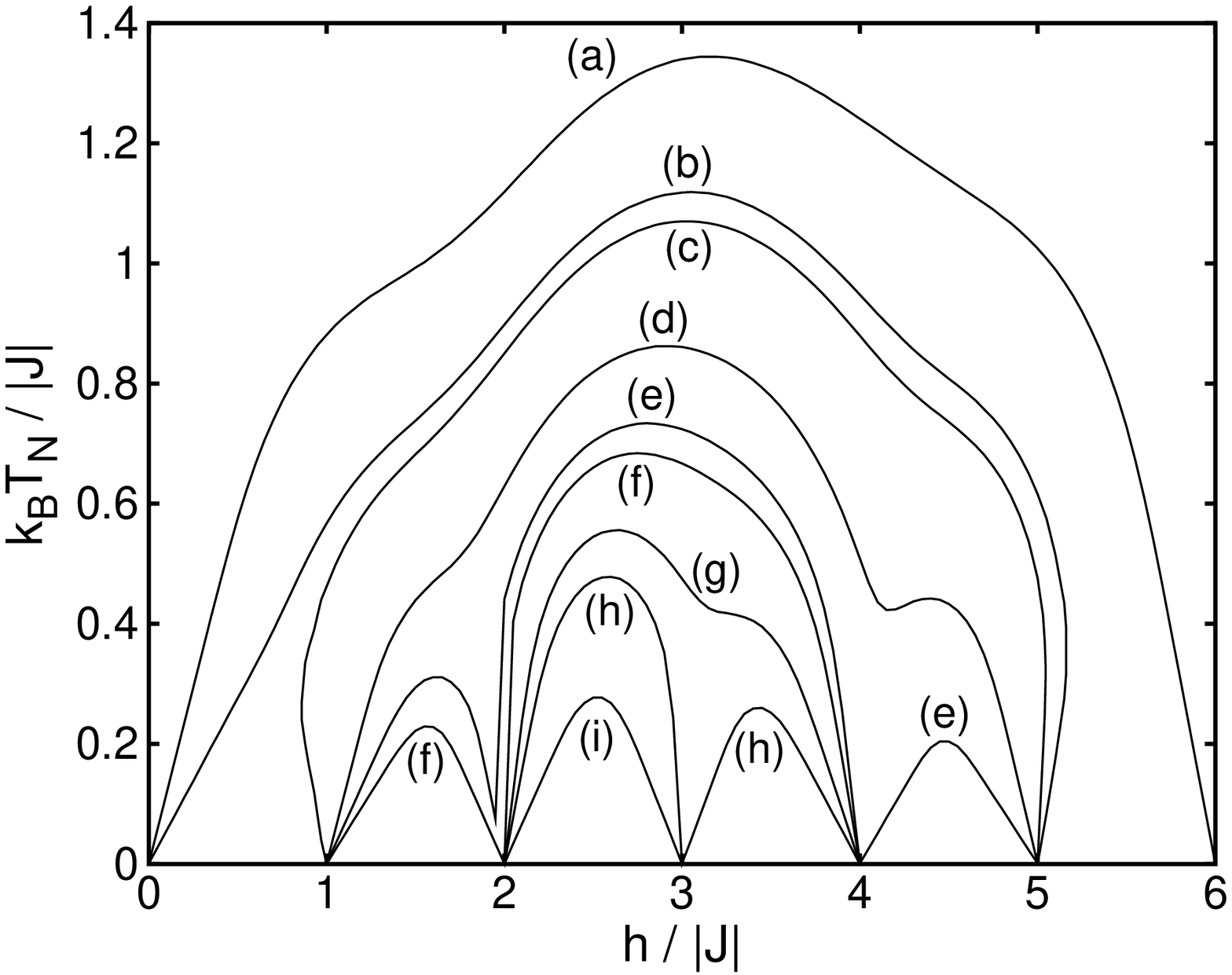}
\caption{Phase boundaries for different values of the concentration $p$: (a) $1$, (b) $0.95$, (c) $0.94$, (d) $0.9$, (e) $0.878$, (f) $0.87$, (g) $0.85$, (h) $0.838$, and (i) 0.81.}
\label{fig:PD}
\end{figure}

\begin{figure}
\centering
    \subfigure[$p=0.95$]{\includegraphics[scale=0.35]{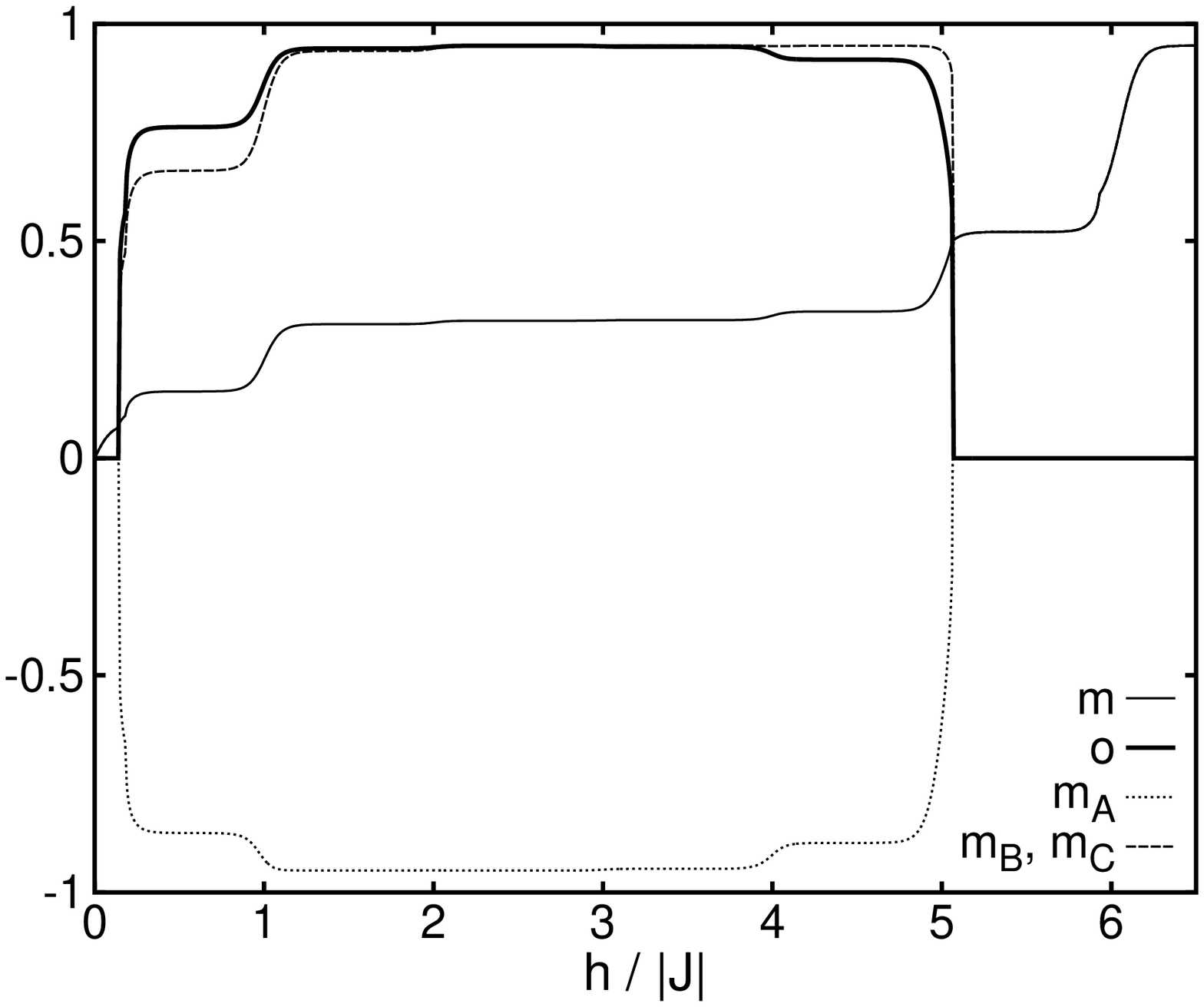}\label{fig:hmp095}}
    \subfigure[$p=0.94$]{\includegraphics[scale=0.35]{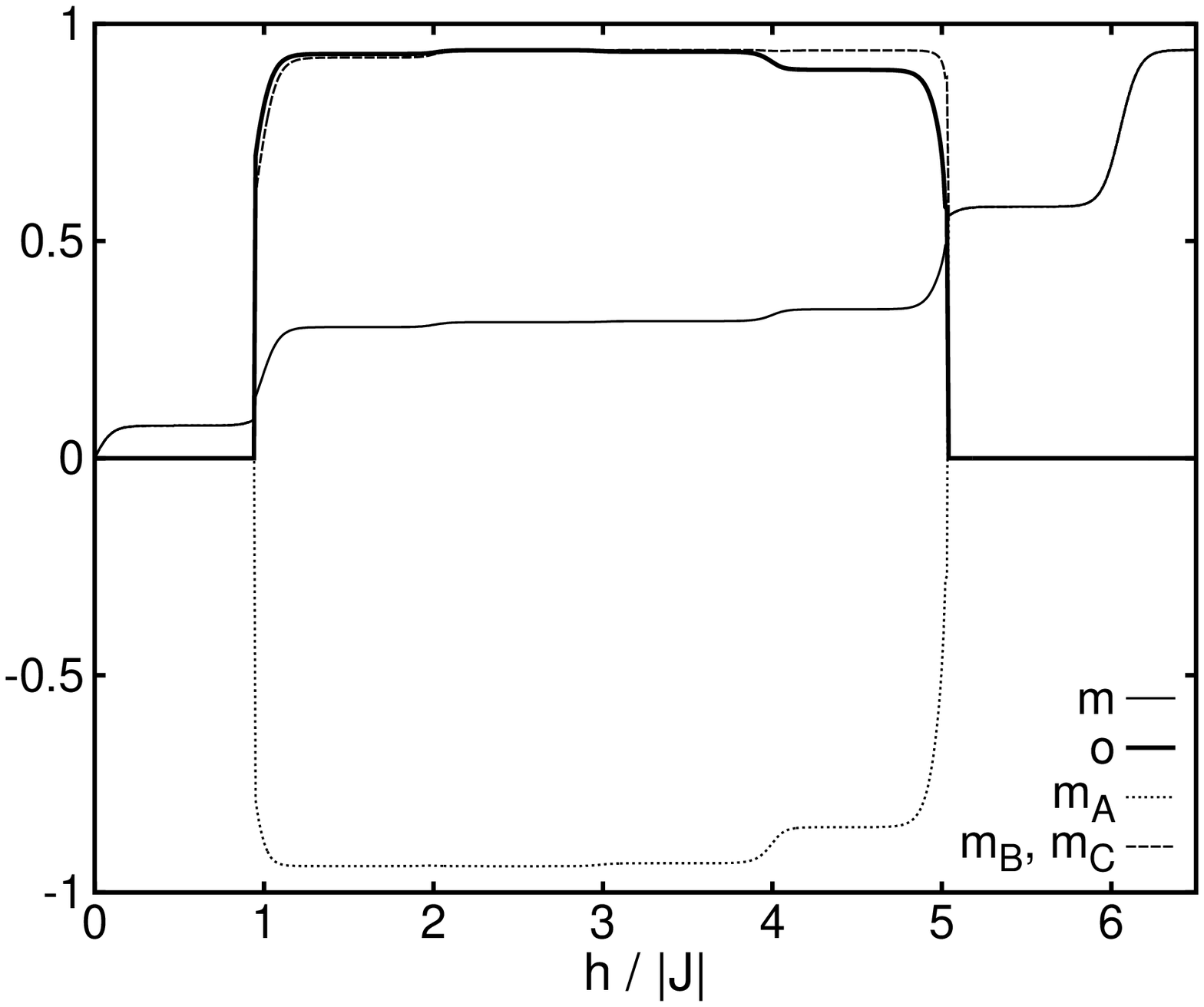}\label{fig:hmp094}}
\caption{Field dependence of the sublattice magnetizations $m_{\mathrm{A}}$, $m_{\mathrm{B}}$, $m_{\mathrm{C}}$, the total magnetization $m$ and the order  parameter $o$, for (a) $p=0.95$ and (b) $p=0.94$ at $k_BT/|J|=0.1$.}
\label{fig:hmp}
\end{figure}

\hspace*{5mm} In order to establish the phase diagrams in the temperature-field-dilution parameter space, we monitor the value of the order parameter $o$. Namely, the transition temperature is set to the value at which the order parameter vanishes. For $p=1$, the phase boundary (curve (a) in Fig.~\ref{fig:PD}) agrees quite well with those obtained in the previous Monte Carlo mean-field, Monte Carlo and renormalization-group studies \cite{metcalf,schick,netz}. The quenched dilution in the present system relieves frustration, however, only locally and thus does not support long-range ordering, in contrast to the selective dilution case \cite{kaya}. Indeed, as evidenced in Fig.~\ref{fig:PD} (see curves (b)-(i)), the increasing dilution lowers the transition temperatures. Furthermore, it shrinks the region of the field values at which the transition can occur. As already signaled by the magnetization behavior shown in Fig.~\ref{fig:m-h}, the shrinking occurs discontinuously and asymmetrically. More specifically, the initial interval of the values at which the transitions can occur $h/|J|\in(0,6)$ for $p=1$ shrinks discontinuously to $(0,5)$ at the concentration $p_{1}=0.96$, then to $(1,5)$ at $p_{2}=0.941$, etc. until $p_6 \equiv p_c=0.802$, below which no long-range order can survive. Such behavior is illustrated in Fig.~\ref{fig:hmp}, where we plot the sublattice magnetizations, the total magnetization, and the order parameter as functions of the applied field for two values of $p=0.95>p_2$ and $p=0.94<p_2$, at $k_BT/|J|=0.1$. 

\begin{figure}
\centering
    \includegraphics[scale=0.5]{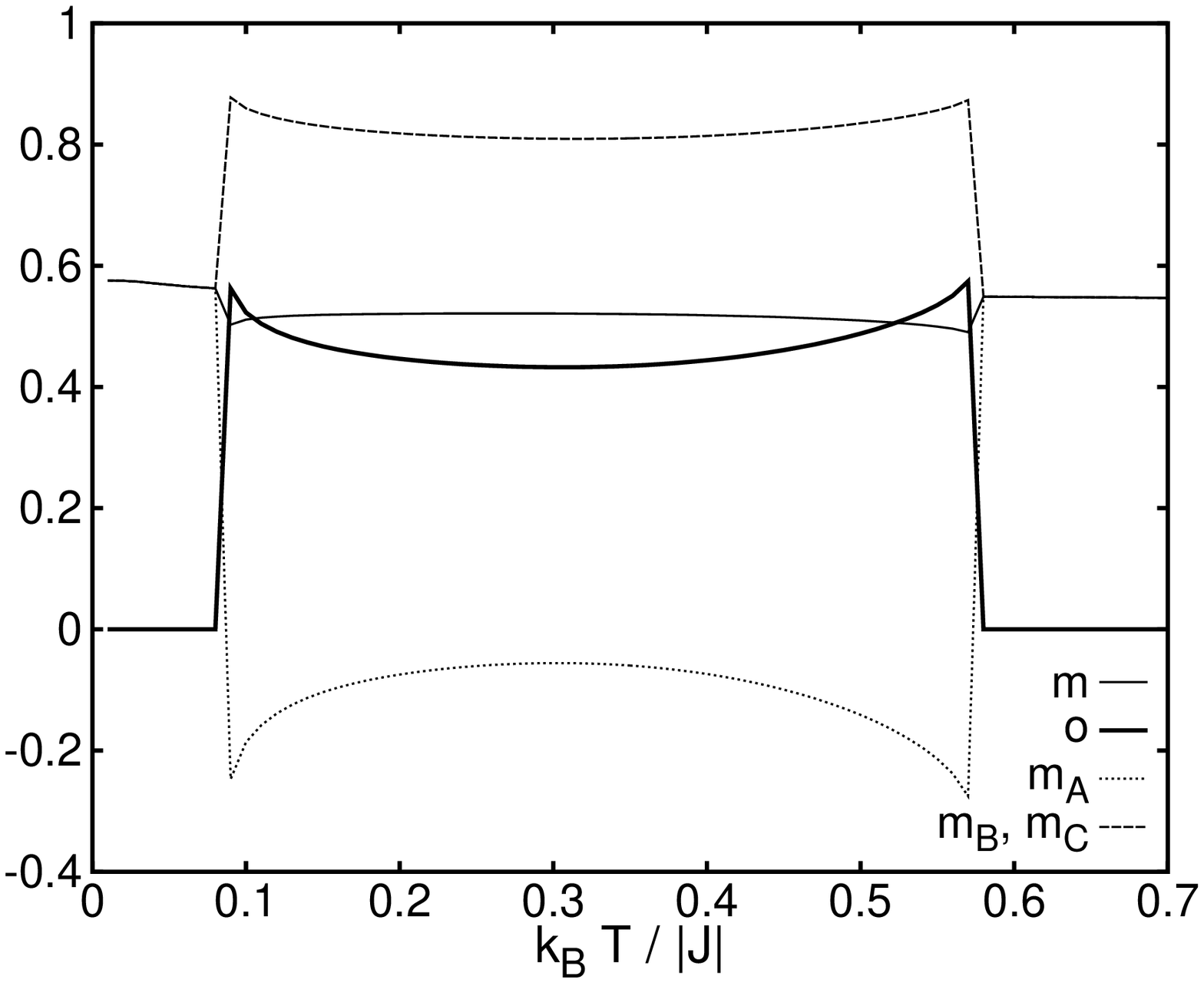}
\caption{Temperature dependence of the sublattice magnetizations $m_{\mathrm{A}}$, $m_{\mathrm{B}}$, $m_{\mathrm{C}}$, the total magnetization $m$, and the order  parameter $o$, for $p=0.95$ (see curve (b) in Fig.~\ref{fig:PD}) at $h/|J|=5.05$.}
\label{fig:reent_t}
\end{figure}

\hspace*{5mm} Owing to the fact that the phase boundaries always terminate at integer values of the field as the ground state is approached, for some values of the field and concentration we can observe a reentrant phenomenon. For example, it occurs for $p=0.95$ (curve (b) in Fig.~\ref{fig:PD}) at $h/|J|=5.05$, as shown in Fig.~\ref{fig:reent_t}. The system passes from the paramagnetic phase at low temperatures to the ferrimagnetic one at higher temperatures and back as temperature is further increased. Such a reentrant behavior can be observed in a low-field region (just below $h/|J|=1$) at concentrations $p\in(0.932,0.941)$ and in a high-field region (just above $h/|J|=5$) at concentrations $p\in(0.934,0.960)$. Thus for $p\in(0.934,0.941)$ the reentrant phenomena appear at both ends of the boundary curve.% \newline\indent

\begin{figure}
\centering
    \includegraphics[scale=0.5]{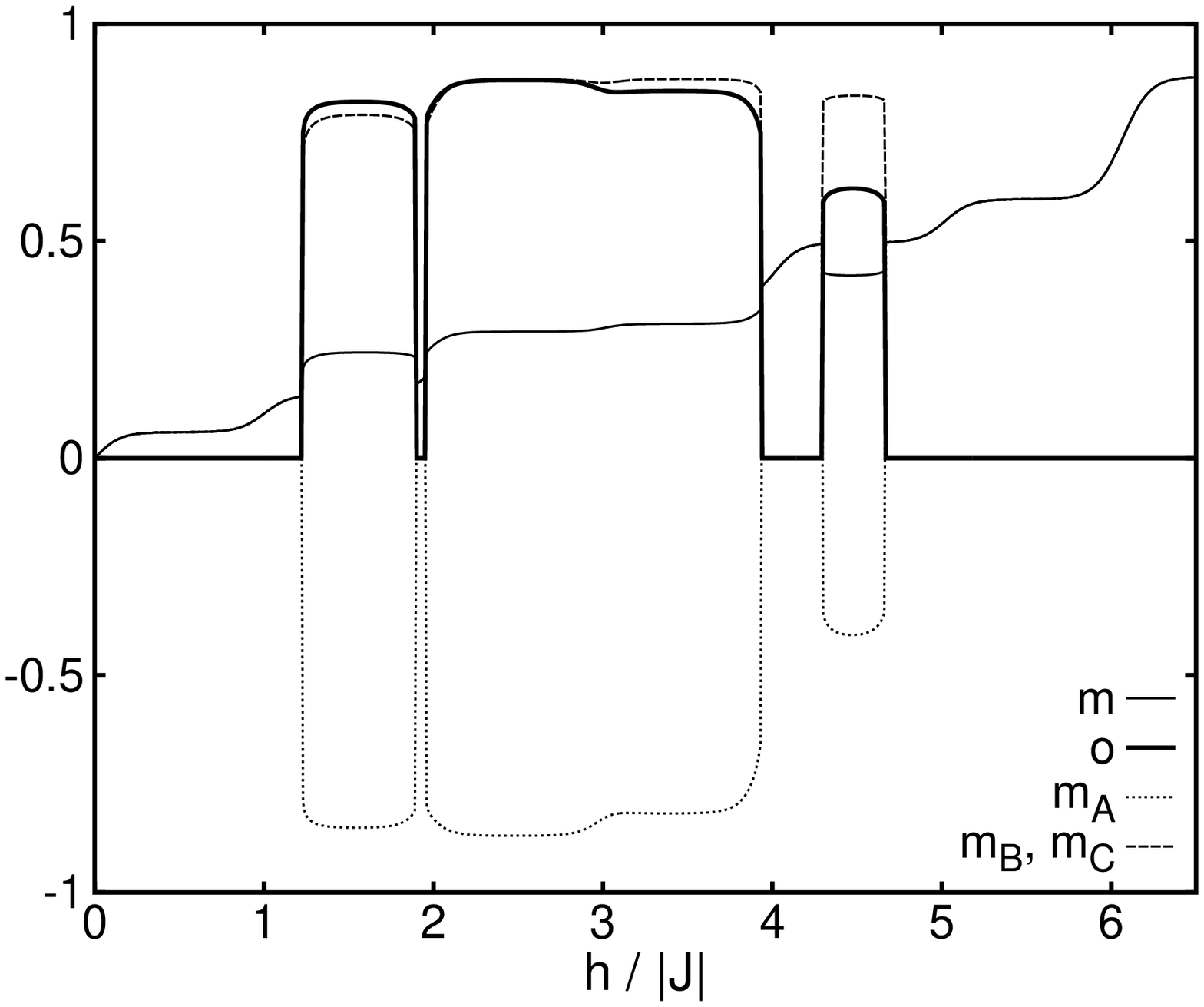}
\caption{Field dependence of the sublattice magnetizations $m_{\mathrm{A}}$, $m_{\mathrm{B}}$, $m_{\mathrm{C}}$, the total magnetization $m$, and the order  parameter $o$, for $p=0.878$ (see curve (e) in Fig.~\ref{fig:PD}) at $k_BT/|J|=0.15$.}
\label{fig:reent_h}
\end{figure}

\begin{figure}
\centering
    \includegraphics[scale=0.5]{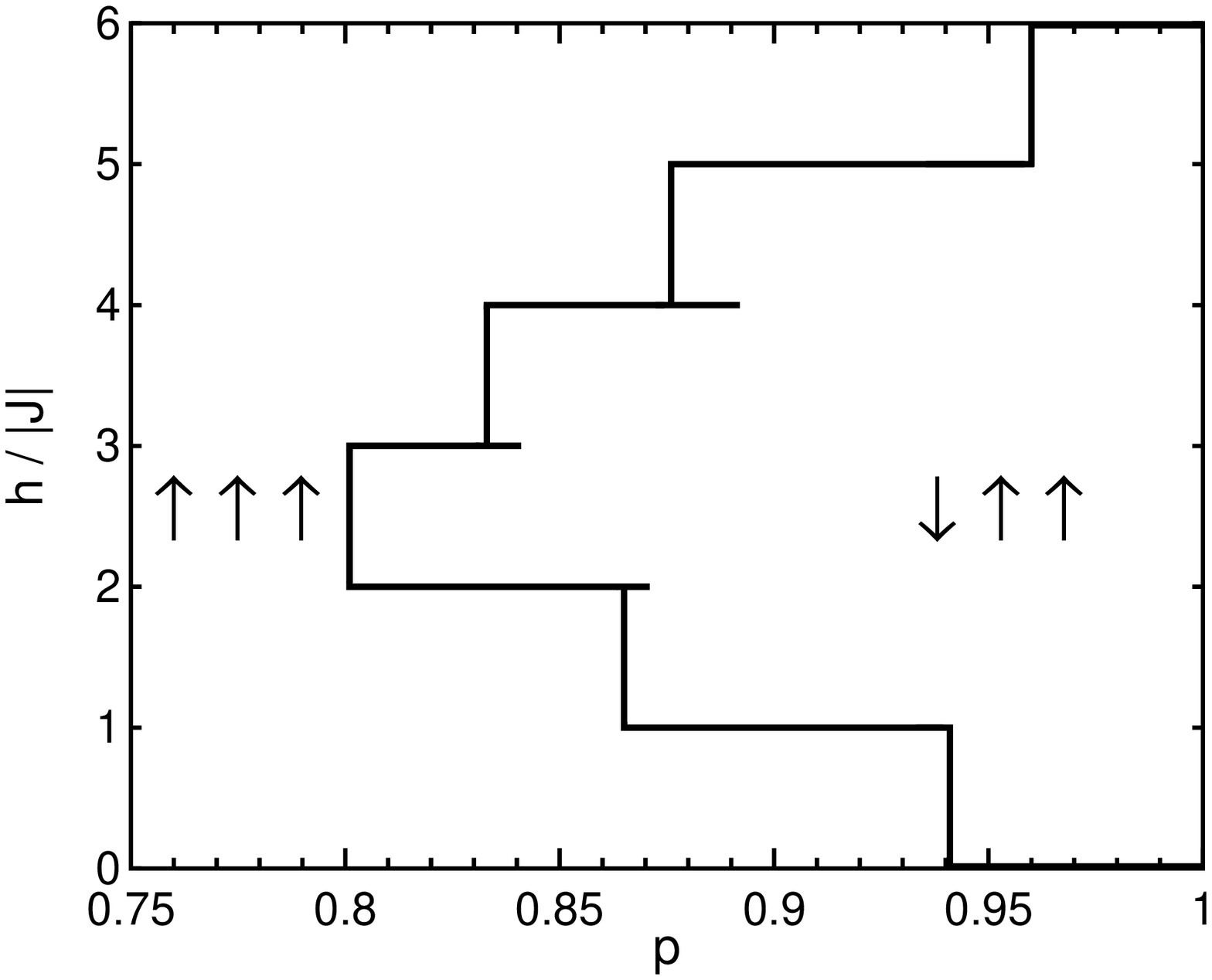}
\caption{Phase diagram in the $h/|J|-p$ parameter space near the ground state at $k_BT/|J|=10^{-6}$.}
\label{fig:GS}
\end{figure}

\hspace*{5mm} In Fig.~\ref{fig:PD} we can also observe that as dilution is increased the boundary curves start to develop a valley. First it appears for $p=0.904$ at the fields just above $h/|J|=4$ (see curve (d)) and gets deeper as dilution increases (curve (e)) until it disappears at $p=0.876$ and the phase transition field interval shrinks from $(1,5)$ to $(1,4)$ (curve (f)). At the same time, another valley emerges at the fields just below $h/|J|=2$ and persists from $p=0.883$ down to $p=0.865$, below which the phase transition field interval shrinks again from $(1,4)$ to $(2,4)$. Eventually, below $p=0.848$ yet another valley appears close to $h/|J|=3$ (curves (g),(h)) until it disappears at $p=0.833$ and the phase boundary becomes limited to the fields $h/|J|\in(2,3)$ (curve (i)). The presence of multiple valleys in the same $(k_BT_c/|J|,h/|J|)$ phase boundary for some values of $p$ enables us to observe reentrant phenomena by varying the field at a fixed temperature. In particular, a typical pure system behavior of passing from the paramagnetic phase at low fields to the ferrimagnetic one at higher fields and back can change to such that, depending on the dilution and temperature values, the system can repeat this cycle twice or even three times, as shown in Fig.~\ref{fig:reent_h}. The latter situation can be observed for $p\in(0.877,0.879)$. As a result, the system displays six percolation thresholds $p_i$, $i=1,\ldots,6$, corresponding to different field intervals, as listed in Table~\ref{tab:percol}. The phase diagram in the $p-h/|J|$ plane close to the ground state (at $k_BT/|J|=10^{-6}$) is shown in Fig.~\ref{fig:GS}. The lines exceeding the edges at $h/|J|=2,3,4$ signify persistence of the reentrant phenomena described above.

\begin{table}
\caption{Percolation thresholds $p_i$ corresponding to different field $h/|J|$ intervals.}
\label{tab:percol}
\centering
\begin{tabular}{lcccccc}
\hline
$h/|J|$  & $(0,1)$  & $(1,2)$ & $(2,3)$  & $(3,4)$ & $(4,5)$  & $(5,6)$ \\
\hline
$p_i$ &  $p_{2}=0.941$ & $p_{4}=0.865$  &  $p_{6}\equiv p_c=0.802$ & $p_{5}=0.833$  & $p_{3}=0.876$ &  $p_{1}=0.96$   \\
\hline
\end{tabular}
\end{table}

\section{Conclusions}
\hspace*{5mm} We have applied an effective-field theory with correlations to study phase transitions in a geometrically frustrated site-diluted triangular lattice Ising antiferromagnet in a field. For the pure system our approach reproduced the exact solution of no long-range order in zero field \cite{wannier,hout} and in a finite field yielded a good agreement with the results obtained by more accurate approaches, such as Monte Carlo simulations \cite{metcalf} and renormalization-group theory \cite{schick}. To our best knowledge, we are not aware of any attempts to establish phase diagrams of the present diluted system. However, our results presented in Fig.~\ref{fig:m-h} nicely corroborate the recent Monte Carlo density of states investigations of magnetization processes in such a system \cite{yao}, namely the stepwise splitting of the low-temperature broad $1/3$ magnetization plateau at integer values of the field as a result of dilution. Although in \cite{yao} no order parameter was evaluated and therefore it is not clear which of the jumps in the magnetization are associated with the phase transitions, it is evident that the low-temperature transitions in that study should occur at some integer values between $0 \leq h/|J| \leq 6$, in accordance with our results. Since the peculiar critical behavior with multiple reentrants observed in our study is a consequence of such discontinuities, we speculate that they are not merely artifacts of the used approximation and might be also reproduced in Monte Carlo simulations. Notwithstanding, further theoretical and experimental evidence is desirable to confirm our predictions. Suitable candidates for the experimental studies could be compounds, such as $\mathrm{Ca_3Co_2O_6}$ or $\mathrm{Cs_3CoX_3}$, (X=Cl or Br) \cite{hardy}.

\section*{Acknowledgments}
This work was supported by the Scientific Grant Agency of Ministry of Education of Slovak Republic (Grant No. 1/0431/10).

\end{document}